\documentclass[12pt,a4paper]{article}       

\usepackage{xcolor}
 \usepackage[skins,theorems]{tcolorbox}
\tcbset{highlight math style={enhanced,
  colframe=red,colback=white,arc=0pt,boxrule=1pt}}
  \usepackage[bookmarksopen, bookmarksnumbered, bookmarksopenlevel=2]{hyperref}
  \usepackage{tikz}
  \usepackage{tikz-3dplot}
 \usetikzlibrary{calc}
 \usetikzlibrary{decorations} %
 \usepackage[UKenglish]{babel}
 \usepackage[toc,page]{appendix}
 \usepackage{amsmath}
 \usepackage{amssymb}
 \usepackage{graphicx}
 \usepackage{hhline}
 \usepackage{tikz-feynman}

 \usepackage[bf]{caption}
\usepackage{cite}
\usepackage[vcentermath]{youngtab}
\usepackage{geometry}
\usepackage{slashed}
\usepackage{stackrel}
\usepackage{tikz-cd} 
\usepackage{mathtools}
\usepackage{cancel} 
\usepackage{multirow}
\usepackage[margin=0pt,font=small,labelfont=normalfont,skip=22pt]{subcaption}

\usepackage{empheq}
\usepackage{arydshln}

\newcommand{\bqa}{\begin{eqnarray}}
\newcommand{\eqa}{\end{eqnarray}}

\newenvironment{eqn*}{\begin{equation*}\begin{aligned}}{\end{aligned}\end{equation*}\noindent}
\hypersetup{
    pdftitle={},
    pdfauthor={},
    pdfsubject={}
}
\numberwithin{equation}{section}
\numberwithin{table}{section}

\makeatletter

\definecolor{dgreen}{rgb}{0,0.45,0.2}
\definecolor{dblue}{rgb}{0,0.0,0.5}

\newcommand{\be}{\begin{equation}}
\newcommand{\ee}{\end{equation}}
\newcommand{\beq}{\begin{equation}}
\newcommand{\eeq}{\end{equation}}
\newcommand{\ba}{\begin{aligned}}
\newcommand{\ea}{\end{aligned}}

\newcommand{\bea}{\begin{eqnarray}}
\newcommand{\eea}{\end{eqnarray}}

\newcommand{\cN}{\mathcal{N}}

\newcommand{\cF}{\mathcal{F}}

\newcommand\bi{\begin{itemize}}
\newcommand\ei{\end{itemize}}

\def\unit{{1\kern-.65ex {\rm l}}}
\def\1{{1\kern-.65ex {\rm l}}}

\def\dd{{\mathrm{d}}}

\newcount\hour \newcount\minute
\hour=\time \divide \hour by 60
\minute=\time
\def\now{%
\ifnum \hour<13
  \ifnum \hour=0 \advance \hour by 12 \number\hour:\else \number\hour:\fi%
     \ifnum \minute<10 0\fi%
     \number\minute%
\ A.M.%
\else \advance \hour by -12 \number\hour:%
  \ifnum \minute<10 0\fi%
  \number\minute%
  \ P.M.%
\fi%
}

\makeatother

\begin{document}

\begin{titlepage}
\begin{center}
\rightline{\small }

\vskip 15 mm

{\large \bf
Bounds on Field Range for Slowly Varying Positive Potentials
} 
\vskip 11 mm

Damian van de Heisteeg,$^{1}$ Cumrun Vafa,$^{2}$ Max Wiesner,$^{1,2}$ David H. Wu$^{2}$

\vskip 11 mm
\small ${}^{1}$ 
{\it Center of Mathematical Sciences and Applications, Harvard University,\\ Cambridge, MA 02138, USA}  \\[3 mm]
\small ${}^{2}$ 
{\it Jefferson Physical Laboratory, Harvard University, Cambridge, MA 02138, USA}

\end{center}
\vskip 17mm

\begin{abstract}
In the context of quantum gravitational systems,  we place bounds on regions in field space with slowly varying positive potentials.  Using the fact that $V<\Lambda_s^2$, where $\Lambda_s(\phi)$ is the species scale, and the emergent string conjecture, we show this places a bound on the maximum diameter of such regions in field space: $\Delta \phi \leq a \ {\rm{log}}(1/V) +b$ in Planck units, where $a\leq \sqrt{(d-1)(d-2)}$, and $b$ is an $\mathcal{O}(1)$ number and expected to be negative.  The coefficient of the logarithmic term has previously been derived using TCC, providing further confirmation.  For type II string flux compactifications on Calabi--Yau threefolds, using the recent results on the moduli dependence of the species scale, we can check the above relation and determine the constant $b$, which we verify is $\mathcal{O}(1)$ and negative in all the examples we studied.
\end{abstract}

\vfill
\end{titlepage}

\newpage

\tableofcontents

\setcounter{page}{1}

\section{Introduction}
The species scale, $\Lambda_s$, introduced in \cite{Dvali:2007hz,Dvali:2009ks,Dvali:2010vm,Dvali:2012uq} is an  effective UV cutoff in theories of quantum gravity capturing the number, $N$, of light degrees of freedom in a quantum theory of gravity via $N\sim \Lambda_s^{-(d-2)}$. In the presence of light modes, $\phi$, the species scale can depend on these fields.  In particular in a recent work \cite{vandeHeisteeg:2022btw} we showed how one can use topological string amplitudes at one loop to compute $\Lambda_s(\phi)$ for type II string theories compactified on Calabi--Yau threefolds.
In this paper we use this to place bounds on the field range for theories which break supersymmetry mildly and lead to slowly varying $V>0$.  Here a mild breaking of supersymmetry means that the breaking does not change the number of light degrees of freedom dramatically.  As we will show in this paper, an example of this is provided by CY threefold flux compactifications.
The basic idea is that in this case, we still can use the $\Lambda_s(\phi)$ computed using topological strings.  For slowly varying fields leading to a quasi-dS space, the idea, that the effective theory does not break down, leads to the requirement that $V<
\Lambda_s^2(\phi)$.  Using the fact that asymptotically the distance conjecture places exponential bounds on the species scale \cite{vandeHeisteeg:2023ubh}, we find that
\begin{equation}\label{eq:intro}
    \Delta \phi \leq \sqrt{(d-1)(d-2)}\cdot {\log(1/V)}+b\,,
\end{equation}
where $b\sim O(1)$;  moreover we find in all the examples we studied $b<0$.  The above bound, except for the constraint on the linear shift term, was previously derived using the Trans-Planckian Censorship Conjecture (TCC) in \cite{Bedroya:2019snp}.  Our argument here is based on the emergent string conjecture~\cite{Lee:2019oct} and consistency of the effective theory, and provides further evidence for the validity of TCC.  Moreover for the first time in this paper we are able to determine the value $b$ in examples of type II compactifications on Calabi--Yau threefolds.  A general bound of the form $\Delta \phi <a \log 1/V$ for $a\sim O(1)$ was previously considered in \cite{Scalisi:2018eaz} to put bounds on inflationary models for which TCC fixes the coefficient $a$. Our results in this paper reproduce the TCC bound from a completely different reasoning and moreover we find the sub-leading correction to the field range encoded in $b$.

The organization of this paper is as follows. In section \ref{sec:speciesandpotentials} we recall the behavior of the species scale and scalar potentials in infinite distance limits, reviewing in particular the bounds put by the TCC \cite{Bedroya:2019snp} on the latter. In section \ref{sec:argument} we argue how the TCC bound can be reproduced from species scale arguments and further use this to bound the field range of slowly varying scalar potentials. In section \ref{sec:CY} we apply these principles to bound the field range of scalar potentials arising in flux compactifications of type II string theory on Calabi--Yau threefolds. We conclude in section \ref{sec:conclusions}. Two ancillary notebooks detailing the computations for the quintic and bicubic examples have been attached to the submission.

\section{Species Scale and Scalar Potentials}\label{sec:speciesandpotentials}

In effective theories of gravity with scalar fields $\phi$ an important question concerns the diameter of the scalar field space as this determines the maximal possible field variation $\Delta \phi$. In effective theories that allow for a consistent UV completion to quantum gravity, the scalar field space $\mathcal{M}$ are generically expected to be non-compact \cite{Ooguri:2006in} implying that the diameter of scalar field spaces consistent with quantum gravity is infinite. On the other hand, the distance conjecture \cite{Ooguri:2006in} implies that close to the infinite distance limits a tower of states becomes light as 
\begin{equation}\label{DC}
    \frac{m}{M_{\rm pl}} \sim e^{-\alpha \Delta \phi}\,,
\end{equation}
with $\Delta \phi $ being the distance in field space and $M_{\rm pl}$ the reduced Planck scale.  The parameter $\alpha$ can be further constrained using the emergent string conjecture \cite{Lee:2019oct} which states that the light modes emerging at infinite distance are either KK modes or the excitations of light fundamental strings. With this input, one can then show \cite{Agmon:2022thq}
$${\sqrt{(D-2)\over (D-d)(d-2)}}\geq \alpha \geq \frac{1}{\sqrt{d-2}}\,,$$
where $d$ is the spacetime dimension and $D\leq 11$ is the higher-dimensional theory which one may decompactify to (additional arguments for the lower bound have been given in \cite{Etheredge:2022opl}). The presence of the light tower of states at large distances invalidates the EFT description above some finite energy scale $\Lambda_{\rm EFT}$ such that the part of the field space which allows for a consistent EFT description is expected to have finite diameter. 

%Si

While the distance conjecture clearly excludes infinite field variations $\Delta \phi\rightarrow \infty$ within a consistent EFT description, it does not tell us much about the diameter of the residual field space once all infinite distance tails are removed. In fact even the refined distance conjecture \cite{Palti:2017elp}, stating that the exponential behavior in \eqref{DC} should set in after one Planck length is traversed, does not give much information about the diameter of the residual field space since it does not constrain the overall normalization in \eqref{DC}. This bears the possibility that for large distances the exponential behavior in \eqref{DC} is realized but that the relevant states still have masses well above the Planck scale. In this context it was recently conjectured in  \cite{Rudelius:2023mjy} that in fact there is always a tower of states with mass below $M_{\rm pl}$ which would effectively constrain the diameter of the effective field space. 

\begin{figure}[!t]
\begin{center}
\includegraphics[width=10cm]{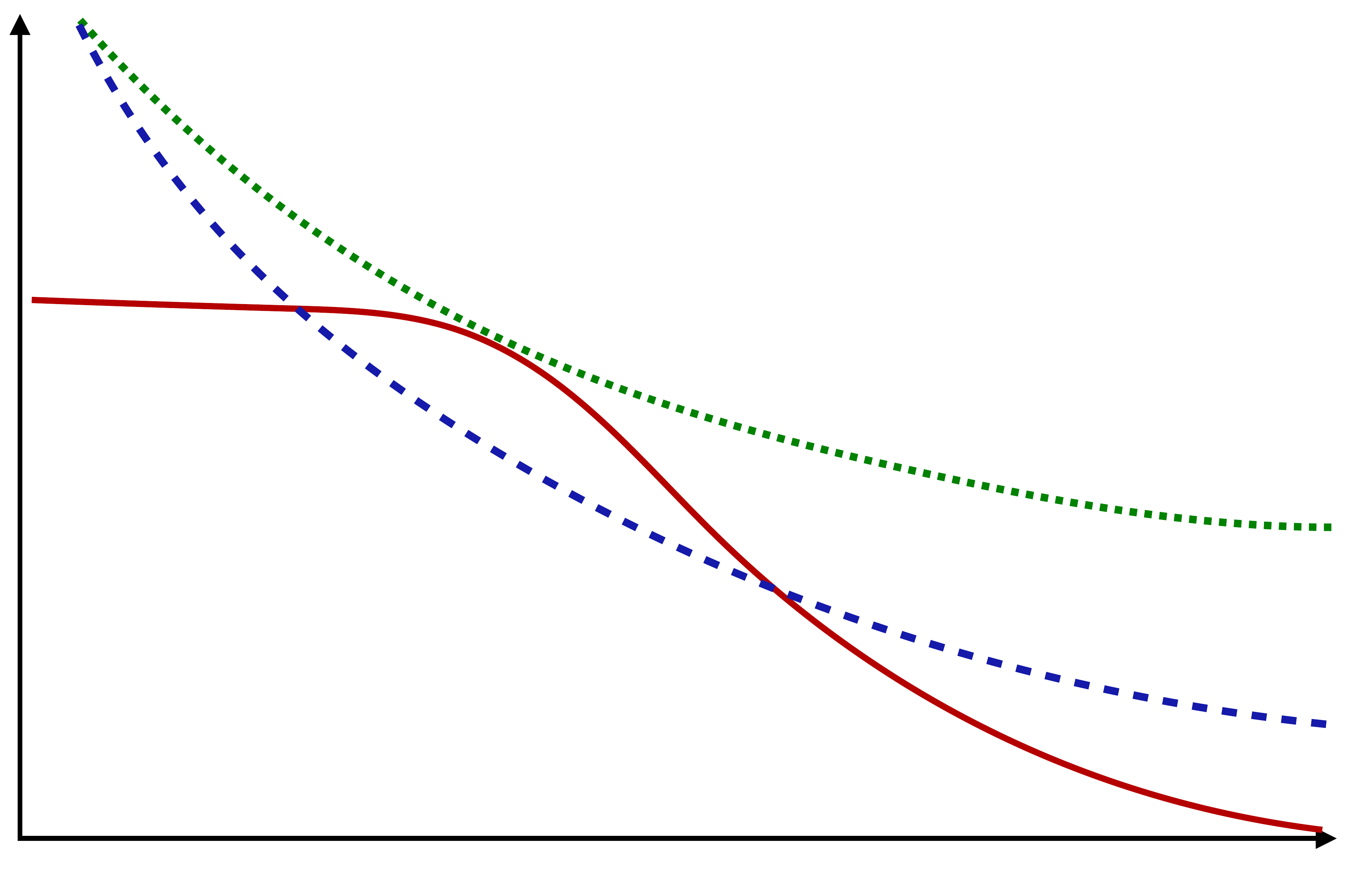}
\end{center}
\begin{picture}(0,0)\vspace*{-1.2cm}
\put(385,37){$\phi$}
\put(90,220){$V(\phi)$}
\put(300,84){\scriptsize{\textcolor{dblue}{${ \exp\left(-\frac{2\Delta \phi}{\sqrt{d-2}}\right)}$}}}
\put(300,120){\scriptsize{\textcolor{dgreen}{${\exp\left(-\frac{2\Delta \phi}{\sqrt{(d-1)(d-2)}}\right)}$}}}
\end{picture}\vspace*{-0.8cm}
\caption{\label{fig:potentialsketch}  An illustration of a potential $V(\phi)$ (solid line) consistent with the constraint imposed by the bound \eqref{eq:boundTCC} (dotted line) and that asymptotically also satisfies the TCC bound \eqref{strongerTCC} for exponentially decaying potentials (dashed line).}
\end{figure}
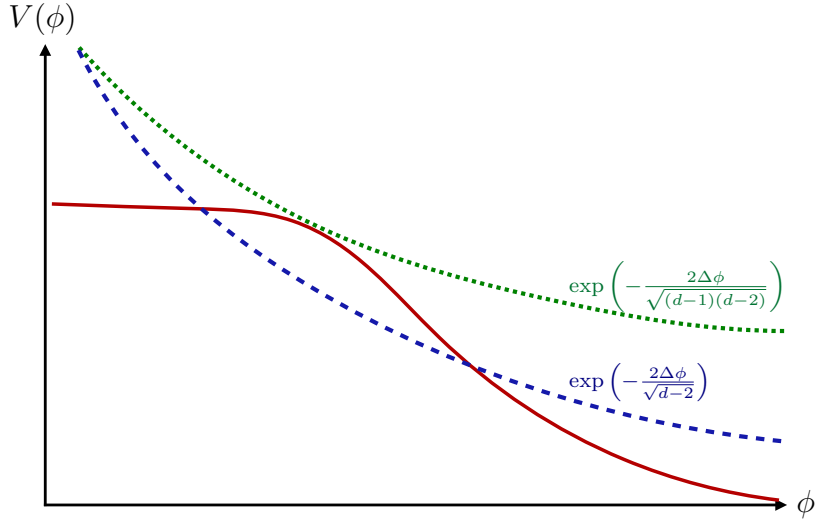

Instead of considering the mass scales of individual towers of states, a more invariant way to capture the moduli-dependence of the EFT cut-off is via the species scale $\Lambda_s$ as this quantity captures the effect of all light towers. The species scale is in general a function of the scalar fields $\phi$ and in certain cases can be evaluated without detailed knowledge about the exact spectrum of light states.  In fact in \cite{vandeHeisteeg:2022btw} we showed that for the case of $\cN=2$ theories in 4d arising from compactifications of type II string theory on Calabi--Yau threefolds, the species scale can be related to the genus-one topological free energy, $F_1$, as $F_1\sim 1/\Lambda_s^2$ and is readily computable.  As noted in \cite{vandeHeisteeg:2023ubh}, the emergent string conjecture \cite{Lee:2019oct} also predicts the species scale in asymptotic regions to fall off exponentially as 
\begin{equation}\label{speciescale}
    \frac{\Lambda_s}{M_{\rm pl}} \sim e^{-\lambda \Delta \phi}\,,
\end{equation}
where 
\begin{equation}\label{boundlambda}
{\sqrt{(D-d)\over (D-2)(d-2)}}\leq \lambda \leq \frac{1}{\sqrt{d-2}}\,. 
\end{equation}
with the highest value corresponding to emergent string limits. Importantly, the lowest value for $\lambda$ is achieved for a decompactification of a single dimensions, i.e.~$D=d+1$, leading to
$$\lambda\geq {1\over {\sqrt{(d-1)(d-2)}}}\,.$$
Apart from the species scale, in theories that allow for a non-trivial scalar potential, there exists a second fundamental scale set by $V(\phi)$. In case of relatively flat potentials (as those required in models of inflation) the relevant scale is the Hubble parameter 
\begin{equation}
 H = \sqrt{V}\,, 
\end{equation}
where we have set $M_{\rm pl}=1$.
In this paper we are mainly interested in studying situations where we indeed have a relatively flat potential. Such scalar potentials have already been severely constrained by the TCC \cite{Bedroya:2019snp}. The TCC asserts that in any consistent theory of quantum gravity any sub-Planckian fluctuation should remain quantum during any cosmological expansion. This leads to a bound on the integral of the Hubble parameter during the expansionary period
\begin{equation}\label{TCCoriginal}
 \int_{t_i}^{t_f} H \leq \log \frac{M_{\rm pl}}{H_f}\,,
\end{equation}
with $H_f$ the Hubble rate at the end of the expansion at time $t_f$. Using the Friedmann equation 
\begin{equation}\label{friedmann}
    (d-1)(d-2) H^2 = \dot{\phi}^2 +2V\,,
\end{equation}
where $\dot{\phi}$ is the time derivative of the rolling scalar field, for $V>0$ one finds
\begin{equation}
    \frac{H}{|\dot{\phi}|}\geq \frac{1}{(d-1)(d-2)}\,. 
\end{equation}
Using that by \eqref{friedmann} $V$ is bounded by $H^2$ one concludes that if a distance $\Delta \phi$ is traversed in field space, in regions where the scalar potential is monotonic (say decreasing), $V$ (in Planck units) should satisfy the bound \cite{Bedroya:2019snp}
\begin{equation}\label{eq:boundTCC}
    V\leq A \ {\rm exp}\Bigg[{-2\Delta \phi \over \sqrt{(d-1)(d-2)}}\Bigg]\,.
\end{equation}
This bound implies that any scalar potential can only stay approximately flat for a finite range in the scalar field space since for sufficiently large distances the scalar potential needs to decay at least exponentially in the canonically normalized scalar field. Its validity has been confirmed in a large class of explicit type II compactifications in \cite{Andriot:2020lea,Andriot:2022xjh}. Notice that the TCC does not explain what forces the scalar potential into this exponential behavior. In this note, by using the species scale, we provide an alternative explanation for the exponential bound in \eqref{eq:boundTCC}. Moreover we find that the maximum range in field space before the exponential behavior sets in occurs for decompactification of a single extra dimension. We show this by connecting the factor of $2/\sqrt{(d-1)(d-2)}$ in \eqref{eq:boundTCC}, coming from the TCC, with the minimal exponent possible for the asymptotic behaviour of the species bound \eqref{boundlambda}. Notice that this does not necessarily imply that the asymptotic shape of the potential is given by this exponent.  Indeed if we assume $V$ has an exponential profile, as one would expect to be the case for large enough field values of $\phi$, $V\sim {\rm exp}(-\beta \phi)$, it was shown in \cite{Bedroya:2019snp} that the TCC implies
\begin{equation}\label{strongerTCC}
    \beta\geq {2\over \sqrt{(d-2)}} \,.
\end{equation}
This has also been argued for in \cite{Rudelius:2022gbz}. In other words TCC would be compatible with a potential which fits the profile in figure~\ref{fig:potentialsketch}.

\section{The Main Argument}\label{sec:argument}
To argue for the bound \eqref{eq:boundTCC} in this paper we make use of the species scale. Recall that the species scale gives a measure for the number, $N$, of light degrees of freedom in an effective theory of gravity via 
\begin{equation}\label{def:speciesscale}
    \Lambda_s =\frac{M_{\rm pl}}{N^{\frac{1}{d-2}}}\,. 
\end{equation}
Let us consider a situation in which we have an approximately flat, positive potential leading to a quasi-dS space. For this dS space the radius of the Hubble horizon is set by the scalar potential 
\begin{equation}\label{radiusHubble}
    r_H \sim \frac{1}{H} = \frac{M_{\rm pl}^{(d-2)/2}}{\sqrt V}\,,
\end{equation}
leading to the Gibbons-Hawking dS entropy \cite{PhysRevD.15.2738}
\begin{equation}
    S_{\rm dS} \sim (r_H M_{\rm pl})^{d-2}\,. 
\end{equation}
Even though in general it is not clear what are the microstates accounting for this entropy, we know that the entropy should \emph{at least} account for the light states in the theory which are counted by $N$. Therefore we find the bound 
\begin{equation}
    N < S_{\rm dS}\,,
\end{equation}
which using \eqref{def:speciesscale} and \eqref{radiusHubble} leads to 
\begin{equation}\label{lambdavsV}
    \Lambda_s \geq \frac{\sqrt{V}}{M_{\rm pl}^{(d-2)/2}}\,. 
\end{equation}
Since the species scale can also be associated to the radius $r_{\rm min}$ of the horizon of the smallest black hole describable within the EFT via $\Lambda_s = r_{\rm min}^{-1}$, the above bound can equivalently be obtained by requiring that this smallest black hole fits within the dS horizon.   

Notice that the bound \eqref{lambdavsV} is also valid in case we have exponentially decaying potentials with $V \sim \exp(-\beta \phi)$ because the Hubble parameter $H$ and $V$ are still related via the Friedmann equation \eqref{friedmann}. Since $H$ defines a horizon radius, we still have the bound \eqref{lambdavsV} by requiring that the horizon of the smallest black hole should be smaller than the cosmological horizon set by $H$. 

Let us mention that the bound \eqref{lambdavsV} has been previously discussed in \cite{Hebecker:2018vxz} and used in \cite{Scalisi:2018eaz} to constrain the maximal field excursion for inflationary models, which was motivated by the requirement that in a consistent EFT the Hubble scale should remain below the quantum gravity cut-off. Identifying the quantum gravity cut-off with the species scale, our previous discussion shows that this is consistent with dS entropy considerations. Using the asymptotic behavior of the species scale as in \eqref{speciescale} the bound \eqref{lambdavsV} was used in \cite{Scalisi:2018eaz} to put an asymptotic bound on $\Delta \phi$ given by 
\begin{equation}
    \Delta \phi \leq \frac{1}{\lambda} \log\frac{M_{\rm pl}}{H}\,. 
\end{equation}
The bound on the field range in this note is in spirit close to that argument, except that in this note we can sharpen this result by finding the actual field range and not just the parametric behavior since, as discussed in the next section, in certain cases we know the explicit form of the species scale also in the interior of field space. In addition, this also allows us to fix the constant coefficients in the expression for the field range $\Delta \phi$. 

As reviewed in the previous section, asymptotically the species scale behaves as 
\begin{equation}\label{asymplambda}
  \frac{\Lambda_{s}}{M_{\rm pl}} = A \, e^{-\lambda \phi}\,,
\end{equation}
with $\lambda$ constrained by the emergent string conjecture as in \eqref{boundlambda}. Notice that since $\lambda\geq\frac{1}{\sqrt{(d-1)(d-2)}}$ the bound \eqref{lambdavsV} leads to \eqref{eq:boundTCC}. The species scale constraint together with the emergent string conjecture hence reproduces the bound imposed by TCC \cite{Bedroya:2019snp}! We can further bound the range, $\Delta \phi_{\rm flat}$, in scalar field  space over which the potential is approximately flat with value of order $V_0$ by 
\begin{equation}
    \Delta \phi_{\rm flat} \leq - \sqrt{(d-1)(d-2)} \log \frac{V_0}{M_{\rm pl}^d} + \log A +\dots\,,
\end{equation}
where the $\dots$ denote additional terms that are suppressed for $V_0 \ll M_{\rm pl}^d$. Let us stress that for large $A$ this would allow the effective field range to possibly be very large. This can be understood by noticing that the species scale is bounded by $\Lambda_s<M_{\rm pl}$: for large $A$ this implies that the exponential behavior can only set in for large values of $\Delta \phi$ allowing for large field ranges with approximately flat scalar potentials. Specifying to flux compactifications of type II string theory on Calabi--Yau threefolds we show in the next section that generically $A\lesssim \mathcal{O}(1)$, thereby severely constraining the maximal field range before the exponential behavior of the scalar potential sets in.

\section{Range of potentials in Calabi--Yau compactifications}\label{sec:CY}
In this section we bound the field ranges for slowly varying scalar potentials arising from Calabi--Yau compactifications of string theory. First we review the species scale definition of \cite{vandeHeisteeg:2022btw}, and argue how it extends to $\mathcal{N}=1$ supergravity theories with scalar potentials. We then consider a set of examples where we explicitly compute the range of the field space for a fixed scalar potential $V=V_0$. 

\subsection{Species scale and fluxes in Type IIB orientifolds}
In \cite{vandeHeisteeg:2022btw} it was argued that the species scale in Type II Calabi--Yau compactifications can be related to the one-loop topological free energy $F_1$. This free energy can be defined from the perspective of the underlying 2d $\mathcal{N}=2$ CFT as the index \cite{Cecotti:1992vy, Bershadsky:1993ta}
\begin{equation}\label{F1worldsheet}
    F_1 = \frac{1}{2} \int_{\mathcal F} \text{Tr}\left[(-1)^F F_L F_R q^{H_0} \bar{q}^{\bar{H}_0}\right]\, ,
\end{equation}
where $F_{L(R)}$ denotes the left-moving (right-moving) fermion number, $\mathcal{F}$ the fundamental domain of SL$(2,\mathbb{Z})$ and $H_0$ the Hamiltonian of the CFT. In \cite{vandeHeisteeg:2022btw} the species scale was related to $F_1$ as (in the following we set $M_{\rm pl}=1$)
\begin{equation}\label{eq:lambdaF1}
    \Lambda_s = \frac{1}{\sqrt{F_1}}\, .
\end{equation}
In \cite{Cribiori:2022nke} this relation was discussed from the perspective of the smallest black holes described by the effective theory. 

The most direct argument given in \cite{vandeHeisteeg:2022btw} for this relation \eqref{eq:lambdaF1} between $\Lambda_s$ and $F_1$ follows from the higher-derivative corrections in the 4d $\mathcal{N}=2$ supergravity theory: $F_1$ appears as coefficient of the higher-derivative term \cite{Bershadsky:1993cx, Antoniadis:1993ze}
\begin{equation}
  S_{\mathcal{N}=2} \supset \int \dd^4x\,  \dd^4\theta \, F_1 \mathcal{W}^2\, ,  
\end{equation}
where $\theta$ denotes the fermionic superspace coordinates, and $\mathcal{W}_{\mu \nu}$ the Weyl superfield. By expanding $\mathcal{W}_{\mu \nu}$ one obtains the $(R^-)^2$ term in the effective action, with $R^-$ the anti-self-dual part of the curvature. The identification of $F_1$ with the species scale $\Lambda_s$ then follows by noting that the coefficient of this term should be given by $M_{\rm pl}^2/\Lambda^2_s$~\cite{vandeHeisteeg:2023ubh}.

Importantly, when breaking the $\mathcal{N}=2$ theory to $\mathcal{N}=1$ by means of fluxes, the topological string amplitudes do not get affected \cite{Vafa:2000wi}. Hence the genus-$g$ topological free energies still determine the coefficients of higher-derivative terms in the effective action. In particular, in the $\mathcal{N}=1$ effective action there are two sorts of higher-derivative corrections \cite{Ooguri:2003qp,Ooguri:2003tt}
\begin{equation}\label{N1flux}
    S_{\mathcal{N}=1} \supset \int \dd^4 x \, \dd^2 \theta \ \cF^{2g} N_i \frac{\partial F_g}{\partial S_i} \, , 
\end{equation}
and
\begin{equation}\label{N1R2}
    S_{\mathcal{N}=1} \supset g\int \dd^4 x \, \dd^2 \theta \ \mathcal{W}^2 \cF^{2(g-1)} F_g\, ,
\end{equation}
where $\mathcal{W}$ now denotes the reduction to an $\mathcal{N}=1$ multiplet, $N_i$ the flux quanta, $S_i$ chiral superfields, and $\cF$ the anti-self-dual part of the graviphoton field strength.  Note in particular that the genus-0 answer from \eqref{N1flux} has been used to compute exact
corrections to $\mathcal{N}=1$ superpotential terms using the breaking of $\mathcal{N}=2$ to $\mathcal{N}=1$ by turning on fluxes \cite{Dijkgraaf:2002dh}.   For us the crucial point is that the fluxes $N_i$ do not appear in equation \eqref{N1R2} and in particular the Weyl-squared term, $\mathcal{W}^2$, is unaffected. This allows us to again relate the species scale $\Lambda_s$ to the genus-one free energy $F_1$. In this sense flux compactifications of Type II string theory on CY3-folds provide an example for which the supersymmetry breaking is mild in the sense that the species scale is not affected considerably by the breaking. 

In order to obtain a closed form for $F_1$, its holomorphic anomaly equation can be integrated \cite{Bershadsky:1993cx}. For Type IIA compactifications this results in an expression that depends only on the K\"ahler moduli. In the $\cN=1$ theory this is still the case where additionally we have to project out all orientifold-even coordinates. The genus-one free energy $F_1$ then reads
\begin{equation}\label{F1}
    F_1 = \frac{1}{2}\left(3+h^{1,1}+\frac{\chi}{12}\right) K^{\tiny{\mathcal{N}=2}}-\frac{1}{2}\log \det G^{{\mathcal{N}=2}}_{i \bar j} + \log |f|^2\, ,
\end{equation}
where $h^{1,1}$ and $\chi$ denote the dimension of the vector multiplet moduli space and Euler characteristic of the Calabi-Yau threefold respectively, $K^{\mathcal{N}=2}$ and $G^{\mathcal{N}=2}_{i\bar{j}}$ the K\"ahler potential and metric of the $\mathcal{N}=2$ parent theory, and the holomorphic ambiguity $f$ can be fixed by matching the asymptotics of $F_1$ at the boundaries of the moduli space.

Finally, let us recall that $F_1$ given by \eqref{F1} is only defined up to an additive constant, as the zero-mode contribution in \eqref{F1worldsheet} diverges. From the perspective of the relation \eqref{eq:lambdaF1} between $\Lambda_s$ and $F_1$ this requires us to introduce an additional coefficient that parametrizes this ambiguity. If we define $F_1$ such that it vanishes at the desert point, this parameter reduces to the number of species $N_{\rm des}$ at the desert point
\begin{equation}\label{eq:lambdaF12}
    \Lambda_s = \frac{1}{\sqrt{F_1+N_{\rm des}}}\, .
\end{equation}
In the following we use this form of the species scale to bound the field range of constant scalar potentials resulting from Calabi-Yau compactifications of Type IIA by using \eqref{lambdavsV}.

\subsection{General considerations}
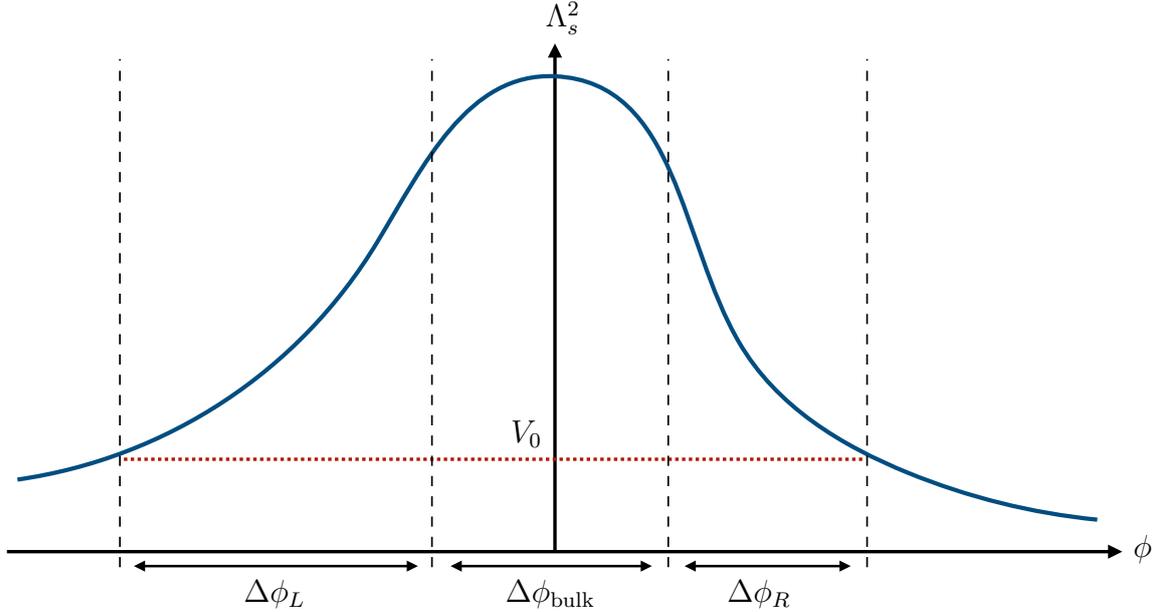
\begin{figure}[!t]
\begin{center}
\includegraphics[width=15cm]{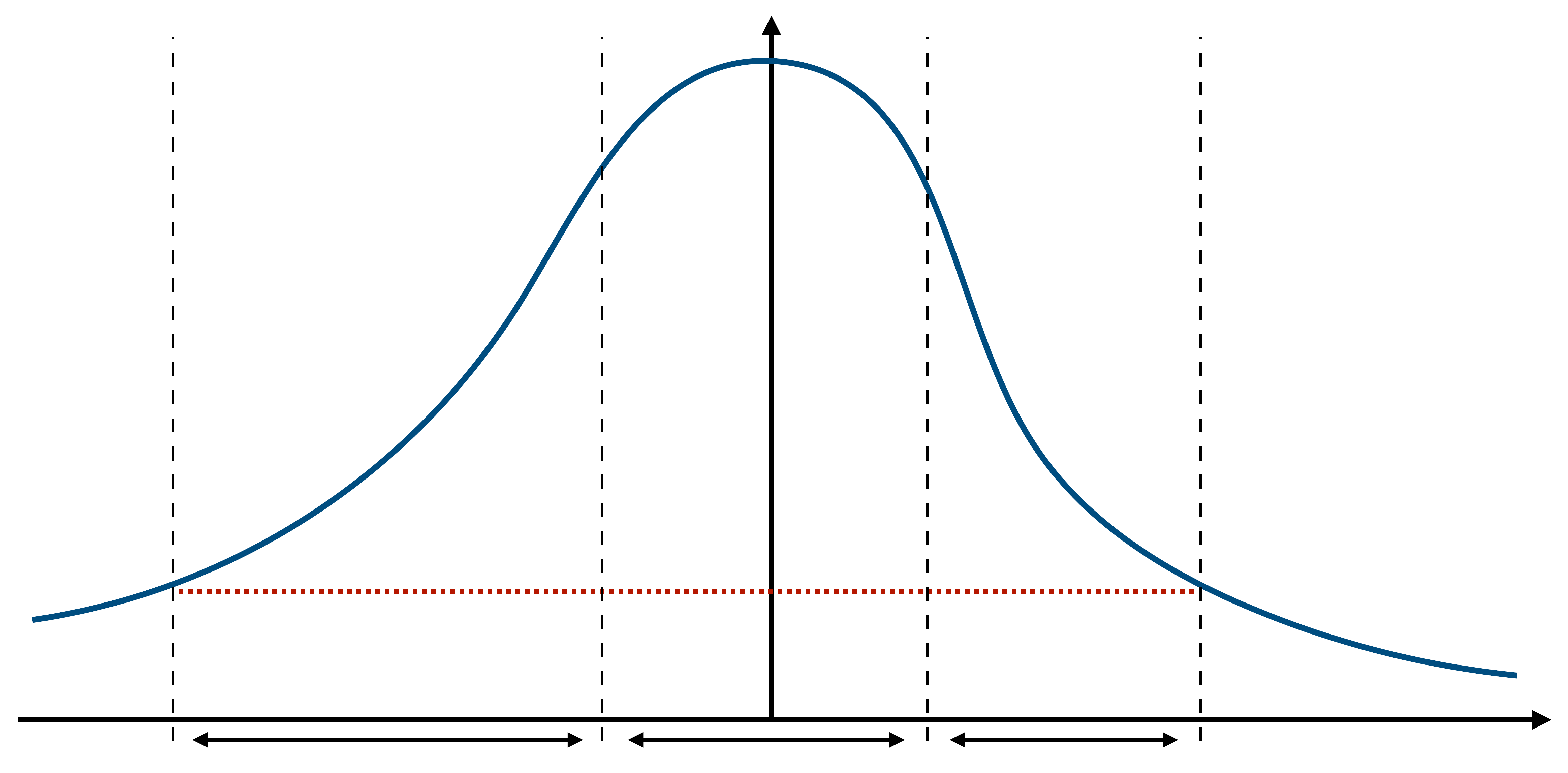}
\end{center}
\begin{picture}(0,0)\vspace*{-1.2cm}
\put(455,40){$\phi$}
\put(235,240){$\Lambda^2_s$}
\put(222,83){$V_0$}
\put(122,23){$\Delta\phi_L$}
\put(220,23){$\Delta\phi_{\rm bulk}$}
\put(303,23){$\Delta\phi_R$}
\end{picture}\vspace*{-0.8cm}
\caption{\label{fig:fieldrange} Sketch of the field range $\Delta\phi$ for a constant scalar potential $V(\phi)=V_0$ (dotted, red) cut off by the species scale $\Lambda_s^2$ going to zero asymptotically (blue, solid). It has been decomposed according to \eqref{eq:fieldrange} into a fixed interior segment $\Delta\phi_{\rm bulk}$ and two tails $\Delta\phi_{L,R}$ probing the infinite distance limits. On the left the species scale goes to zero at a slower exponential rate compared to the right such that the field range extends further into deeper into the left asymptotic regime.}
\end{figure}

In this (and the following) subsection we consider constant potentials $V(\phi)=V_0$ in Calabi-Yau compactifications, and ask ourselves how the bound $V \leq \Lambda_s^2$ precisely restricts their field range. This range often extends across multiple phases of the moduli space, so we will have to carefully examine both the size of the interior of the compact moduli space, as well as the precise asymptotics near its infinite distance boundaries. In this first subsection we approach this problem from a general viewpoint, decomposing the field range into multiple pieces and estimating the contributions coming from each of them, as depicted in figure \ref{fig:fieldrange}.

\subsubsection{Field range of constant potentials}
We denote the field range for a given energy scale $V_0$ by $\Delta\phi(V_0)$. This field range is computed by considering the longest minimal geodesics in the moduli space along which $V_0 \leq \Lambda_s^2$ is satisfied. Concretely, this geodesic has two endpoints that we may move further into infinite distance regimes as we decrease $V_0$. We may therefore break down the range $\Delta\phi(V_0)$ into three contributions\footnote{Notice that for certain moduli spaces there is only one infinite distance limit in the fundamental domain, as happens for instance for the quintic. In that case we fix one of the endpoints at the desert point, meaning that one tail has zero length $\Delta\phi_L=0$, and we only get contributions from the interior phase $\Delta\phi_{\rm bulk}$ and the other endpoint $\Delta\phi_{R}$. }
\begin{equation}\label{eq:fieldrange}
    \Delta\phi(V_0) = \Delta\phi_L(V_0)+\Delta\phi_{\rm bulk}+\Delta\phi_{R}(V_0)\, .
\end{equation}
The middle contribution $\Delta\phi_{\rm bulk}$ comes from the interior of the moduli space; throughout most of this paper we assume $V_0$ to be sufficiently small, so we traverse this phase in its entirety, and hence $\Delta\phi_{\rm bulk}$ represents a constant contribution to the field range. Estimating the size of $\Delta\phi_{\rm bulk}$ is a crucial piece in order to determine when the exponential behavior of the distance conjecture \cite{Ooguri:2006in} sets in.\footnote{ A closely related problem concerns the computation of the diameter of the non-geometric phase of the moduli space: in \cite{Blumenhagen:2018nts,Klawer:2021ltm} this problem has been addressed in examples by calculating the length of geodesics between, e.g., the LG and conifold points without any reference to scalar potentials. Instead our definition of $\Delta \phi_{\rm bulk}$ in terms of the species scale applies to more general situations, as it allows to compute the bulk field range in any direction of the field space even in the absence of LG points. }

The other two contributions $\Delta\phi_L(V_0)$ and $\Delta\phi_R(V_0)$ come from the two endpoints along infinite distance limits where $\Lambda_s^2$ goes to zero. These distances can be computed by considering just the asymptotic behaviors associated to these phases. Taking $\Delta\phi_R$ for definiteness, it behaves as
\begin{equation}\label{eq:deltaphiasymp}
    \Delta\phi_R(V_0) = -\frac{1}{2\lambda_R}\log[V_0] +\frac{1}{2\lambda_R}\log[A_R]\, .
\end{equation}
where we are using that the species scale behaves asymptotically as $\Lambda_s^2 = A_R e^{-2\lambda_R \Delta\phi_R}$. In total the field range may thus be estimated as
\begin{equation}\label{DeltaphiV0}
\Delta\phi(V_0) = -\left(\frac{1}{2\lambda_L}+\frac{1}{2\lambda_R}\right)\log[V_0] +\underbrace{\frac{1}{2\lambda_L}\log[A_L]+\frac{1}{2\lambda_R}\log[A_R]+\Delta\phi_{\rm bulk}}_{=:b}\, ,
\end{equation}
Thus we can maximize the field range for small $V_0$ by selecting limits with the smallest coefficient $\lambda_{L/R}$. From the discussion in the previous section we know that this corresponds to the case that both asymptotic limits to the left and the right correspond to decompactification limits to one dimension higher. The last three terms on the RHS in \eqref{DeltaphiV0} are independent of $V_0$ and can be identified with the constant shift $b$ in \eqref{eq:intro}. The sign of $b$ is hence determined by the value of $\log A_{L/R}$ relative to $\Delta \phi_{\rm bulk}$. From the analysis of \cite{Blumenhagen:2018nts} we expect that the contribution of non-geometric phases to $\Delta \phi_{\rm bulk}$ is of $\mathcal{O}(1)$ in Planck units.\footnote{Reference \cite{Klawer:2021ltm} identifies possible geometries for which $\Delta \phi_{\rm bulk}$ can be made larger than unity in Planck units. However, in the examples studied a hierarchy $\Delta \phi_{\rm bulk}\gg 1$ cannot be achieved. It would be interesting to study the relation between $\Delta \phi_{\rm bulk}$ and the scaling of $A_{L/R}$ in these examples in more detail.} On the other hand, by studying the asymptotics of the species scale we show in the following that typically $A_{L/R}\ll  1$, such that their respective contributions to $b$ overcompensate the positive $\Delta \phi_{\rm bulk}$ leading to a negative sign for $b$. We further confirm this expectation in explicit examples in section~\ref{sec:examples}. 

\subsubsection{Asymptotics of the field range}
While we cannot fix the radius of the interior $\Delta\phi_{\rm bulk}$ on general grounds, we can bound the coefficients appearing in the infinite distance limits $\Delta\phi_{L,R}$ in \eqref{eq:deltaphiasymp}. The coefficients $\lambda_{L,R}$ of the species scale are already constrained by \eqref{boundlambda}. As we will show now, the $\mathcal{O}(1)$ factors $A_{L,R}$ can be bounded as well when considering the asymptotics of $F_1$ in various limits in the large volume regime.

We parametrize this phase by coordinates $t^i=b^i+iv^i$, with the large volume regime corresponding to $v^i\gg 1$. By using the asymptotic behavior of \cite{Bershadsky:1993ta} for $F_1$ in this limit, we can give the behavior of the species scale as
\begin{equation}\label{eq:speciesLV}
    \Lambda_s^2 = \frac{12}{2\pi c_{2,i} v^i} + \mathcal{O}(v^{-2}\log[v])\, ,
\end{equation}
where $c_{2,i}$ are the second Chern class numbers of the Calabi-Yau threefold. We now want to consider sublimits in the large volume regime, where we take a subset of volumes to infinity. To this end we split the coordinates as $t^i=(b^a+iv^a,b^\alpha+iv^\alpha)$, and scale the volumes as
\begin{equation}
    v^a = \hat{v}^a s\, , \qquad v^\alpha = \hat{v}^\alpha\, ,
\end{equation}
where we send $s \to \infty$,  with the coefficients  $\hat{v}^i=(\hat{v}^a,\hat{v}^\alpha)$ fixed. Setting the point where the large volume phase starts to be $s=1$ (which may be achieved by rescaling the $\hat{v}^i$ appropriately), we can compute the distance in terms of $s$ as
\begin{equation}
    \Delta\phi = \sqrt{\frac{n}{2}}\log[s]\, .
\end{equation}
where the integer $n=1,2,3$ depends on the choice of limiting coordinates $v^a$ and the fibration structure of the Calabi--Yau manifold. This allows us to express the species scale \eqref{eq:speciesLV} in terms of the distance as
\begin{equation}
    \Lambda_s^2 = \frac{12}{2\pi c_{2,a}\hat{v}^a} e^{-\sqrt{\frac{2}{n}}\Delta\phi}\, .
\end{equation}
From this expression we can identify the order one coefficient $A$ in \eqref{asymplambda} as 
\begin{equation}
    A = \frac{12}{2\pi c_{2,a}\hat{v}^a}\, ,
\end{equation}
where we note that the sum over second Chern class coefficients $c_{2,a}$ runs only over the volumes $v^a$ which are taken to infinity in our limit. For the case of an emergent string limit ($n=1$) --- which corresponds to a K3-fibration where we send only the volume $v^1$ of the $\mathbb{P}^1$ base to infinity --- we have $c_{2,1}=24$ leading to
\begin{equation}
    A_{\rm ES} = \frac{1}{4\pi \hat{v}^a}\, .
\end{equation}
Identifying $\hat{v}^1$ with the string coupling in the dual heterotic picture, we are led to bounding $\hat{v}^1 \gtrsim 1$ and thus $A_{\rm ES} \lesssim 1/4\pi$. In one-dimensional ($n=3$) and two-dimensional ($n=2$) decompactification limits demanding the volumes to be bounded from below yields similar upper bounds $\hat{v}^a\gtrsim 1$ on $A$. Together with the condition that the second Chern class coefficients are integer quantized, $c_{2,i} \in \mathbb{N}$, this leads to $A\leq\mathcal{O}(1)$. As a consequence of the non-negativity of each of the $c_{2,i}$, in multi-moduli limits the parameter is generically much smaller than one. In the following we show that even for one-parameter models $A$ is sufficiently small such that the constant shift $b$ in \eqref{DeltaphiV0} is negative.

\subsection{Examples}\label{sec:examples}
In this section we use the bound $V \leq \Lambda_s^2$ to determine the range of constant potentials $V(\phi)=V_0$ in examples of Calabi--Yau compactifications, taking the same models as considered in \cite{vandeHeisteeg:2022btw, vandeHeisteeg:2023ubh}. For this field range we consider geodesic paths whose endpoints move further towards infinite distance points as we decrease $V_0$.\footnote{When the fundamental patch of the moduli space has only one infinite distance limit, such as e.g.~the quintic, we fix one of the two endpoints at the desert point.} By using the explicit form \eqref{eq:lambdaF12} of the species scale we are able to give a numerical relation between the distance $\Delta \phi$ and the value of the potential $V_0$ in all examples.

\subsubsection{Example 1: \texorpdfstring{$(K3\times T^2)/\mathbb{Z}_2$}{K3 x T2}}
\begin{figure}[!t]
\begin{center}
\includegraphics[width=10cm]{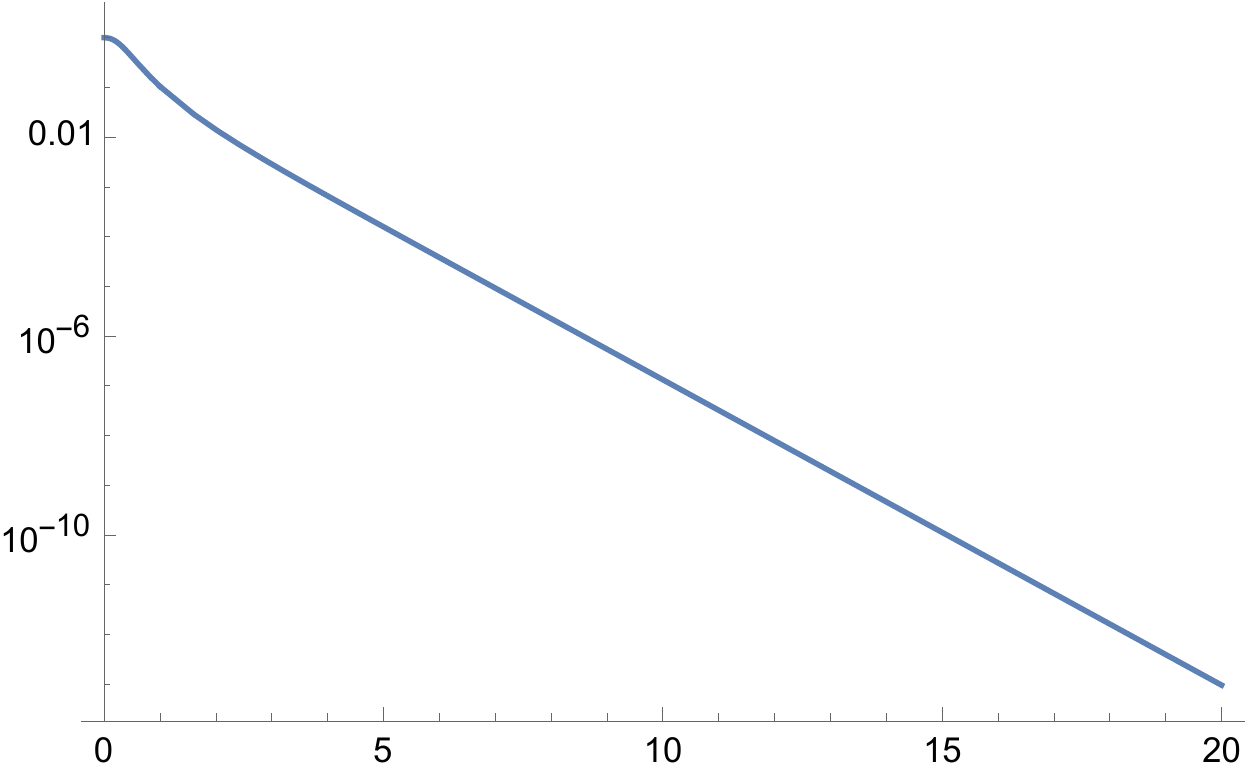}
\end{center}
\begin{picture}(0,0)\vspace*{-1.2cm}
\put(389,38){$\phi$}
\put(117,213){$\Lambda_s^2$}
\end{picture}\vspace*{-0.8cm}
\caption{\label{fig:T2species} Logarithmic plot of the square of the species scale for Enriques Calabi-Yau $(K3\times T^2)/\mathbb{Z}_2$ against the distance $\phi$ in scalar field space. As path in field space we take the desert point $t=\frac{1}{2}(-1+i\sqrt{3})$ as starting point, and move towards $t=i\infty$ while keeping the real part of $t$ -- the axion -- fixed. We see that the asymptotic exponential behavior of $\Lambda_s^2$, corresponding to linear behavior in this logarithmic plot, sets in within roughly one Planck length $\phi \gtrsim 1$. }
\end{figure}
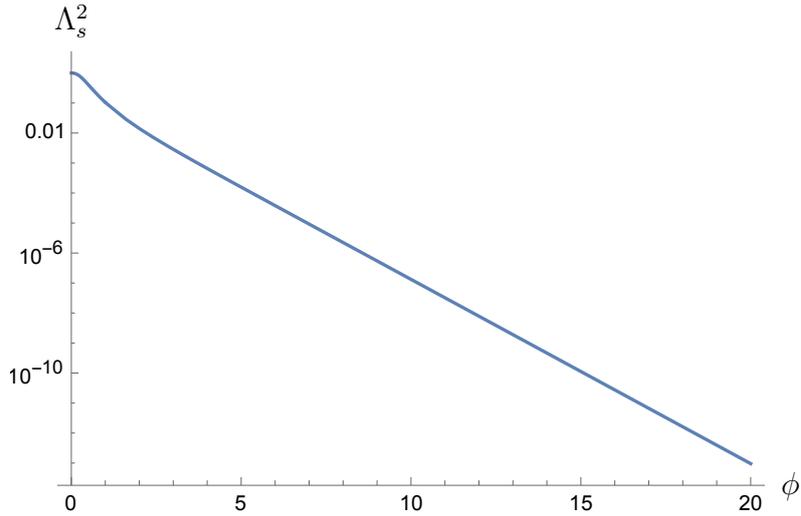

We begin with a compactification of Type IIA on Enriques Calabi-Yau $(K3\times T^2)/\mathbb{Z}_2$. We suppress all dependence on the K3-moduli, considering only the K\"ahler parameter of the two-torus $T^2$. In that case the topological free energy is given by \cite{Bershadsky:1993ta}
\begin{equation}\label{eq:F1T2}
    F_1 = -6\log\left[i(\bar{t}-t)|\eta^2(t)|^2\right]+6\log\left[\frac{3\Gamma(\frac{1}{3})}{16\pi^4}\right]\, ,
\end{equation}
which we normalized such that $F_1(e^{2\pi i /3})=0$. In order to gain intuition for the behavior of the species scale we have provided a plot of $\Lambda_s$ against the field space distance $\phi$ in figure \ref{fig:T2species}.

Using the explicit expression \eqref{eq:F1T2} for $F_1$ --- and thus via  \eqref{eq:lambdaF12} for the species scale $\Lambda_s$ --- we can compute the range of the potential as a function of $V_0$ by requiring $V_0 \leq \Lambda_s^2$. We compute this diameter by considering a path\footnote{Note that we have made the simplifying assumption here that the axion stays fixed for the geodesic of maximal length in our reduced field space. In principle all axion values $-\frac{1}{2}\leq a \leq \frac{1}{2}$ should be considered for the endpoint where $\Lambda_s^2=V_0$. However, as we go further along the infinite distance limit, the contribution from the axionic field displacement becomes exponentially small in the distance, as it is bounded from above by a triangle inequality argument by the length of the segment $-\frac{1}{2}\leq a \leq \frac{1}{2}$ at the endpoint $s=s_{\rm ES}$.}
\begin{equation}
    t(s) = -\frac{1}{2} +i s\, , \qquad \sqrt{3}/2 \leq s \leq s_{\rm ES}\, .
\end{equation}
with $s$ the affine parameter. The starting point is located at the desert point, while the endpoint $s_{\rm ES}$ lies along the emergent string limit $s \gg 1$, defined by where the inequality $V_0\leq \Lambda_s^2$ is saturated. The distance between these two endpoints is computed with the standard hyperbolic metric $K_{ss} = 1/2s^2$ on the moduli space, which yields
\begin{equation}
    \Delta \phi = \frac{1}{\sqrt{2}} \left(\log[s_{\rm ES}]-\log[\sqrt{3}/2]\right)\, .
\end{equation}
In order to give a precise characterization of the relation between the range $\Delta\phi$ and the gap $V_0$ we expand $\Lambda_s$ for large imaginary $t=-\tfrac{1}{2}+is$. In this limit the exact expression \eqref{eq:F1T2} can be approximated by
\begin{equation}\label{eq:speciesT2}
    \Lambda_s^2 = \frac{1}{2\pi s%-6\log[s]+6\log\left[\tfrac{3\Gamma(\tfrac{1}{3})^6}{32\pi^4}\right]+1} +\mathcal{O}(e^{-2\pi s})
    }+\mathcal{O}(s^{-2}\log[s])\, .
\end{equation}
This approximation suffices to determine the relation between $\Delta\phi$ and $V_0$ for small $V_0$ as
\begin{equation}
\begin{aligned}
    \Delta\phi &= -\frac{1}{\sqrt{2}} \log[V_0] -\frac{1}{\sqrt{2}}\log[\sqrt{3}\pi] \simeq  -\frac{1}{\sqrt{2}} \log[V_0] -1.198\, .
\end{aligned}
\end{equation}
The first term gives the expected exponential relation between $V_0$ and $\Delta\phi$, with $1/\sqrt{2}$ the coefficient of the emergent string limit. In the light of \eqref{DeltaphiV0} the negative second term can be attributed to the sum of the $\frac{1}{\lambda} \log A = \tfrac{1}{\sqrt{2}}\log \tfrac{1}{2\pi} \approx -1.3$ and $\Delta \phi_{\rm bulk}\approx 0.1 $ measuring the distance between the desert point and $t=-\tfrac{1}{2}+i$ where the exponential behavior of the species scale sets in.

\subsubsection{Example 2: Mirror Quintic \texorpdfstring{$X_5(1^5)$}{X5}}
\begin{figure}[!t]
\begin{subfigure}{0.49\textwidth}
\begin{center}
\includegraphics[width=0.82\textwidth]{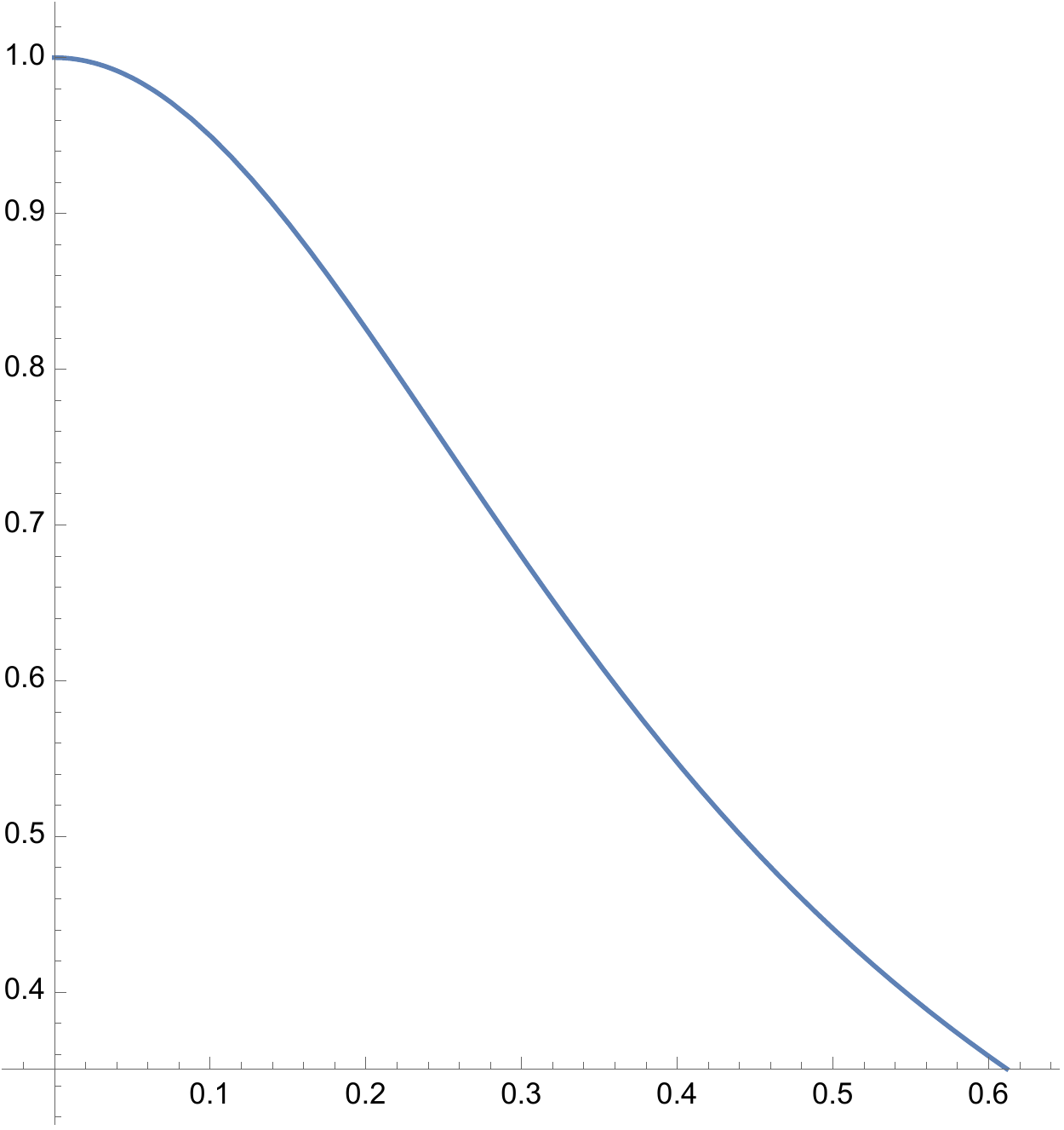}
\end{center}
\begin{picture}(0,0)
\put(222,36){$\phi$}
\put(25,240){$\Lambda^2_s$}
\end{picture}
\vspace*{-40pt}\caption{LG phase.\label{fig:LGphase}}
\end{subfigure}
\hspace{2pt}
\begin{subfigure}{0.49\textwidth}
\begin{center}
\includegraphics[width=0.9\textwidth]{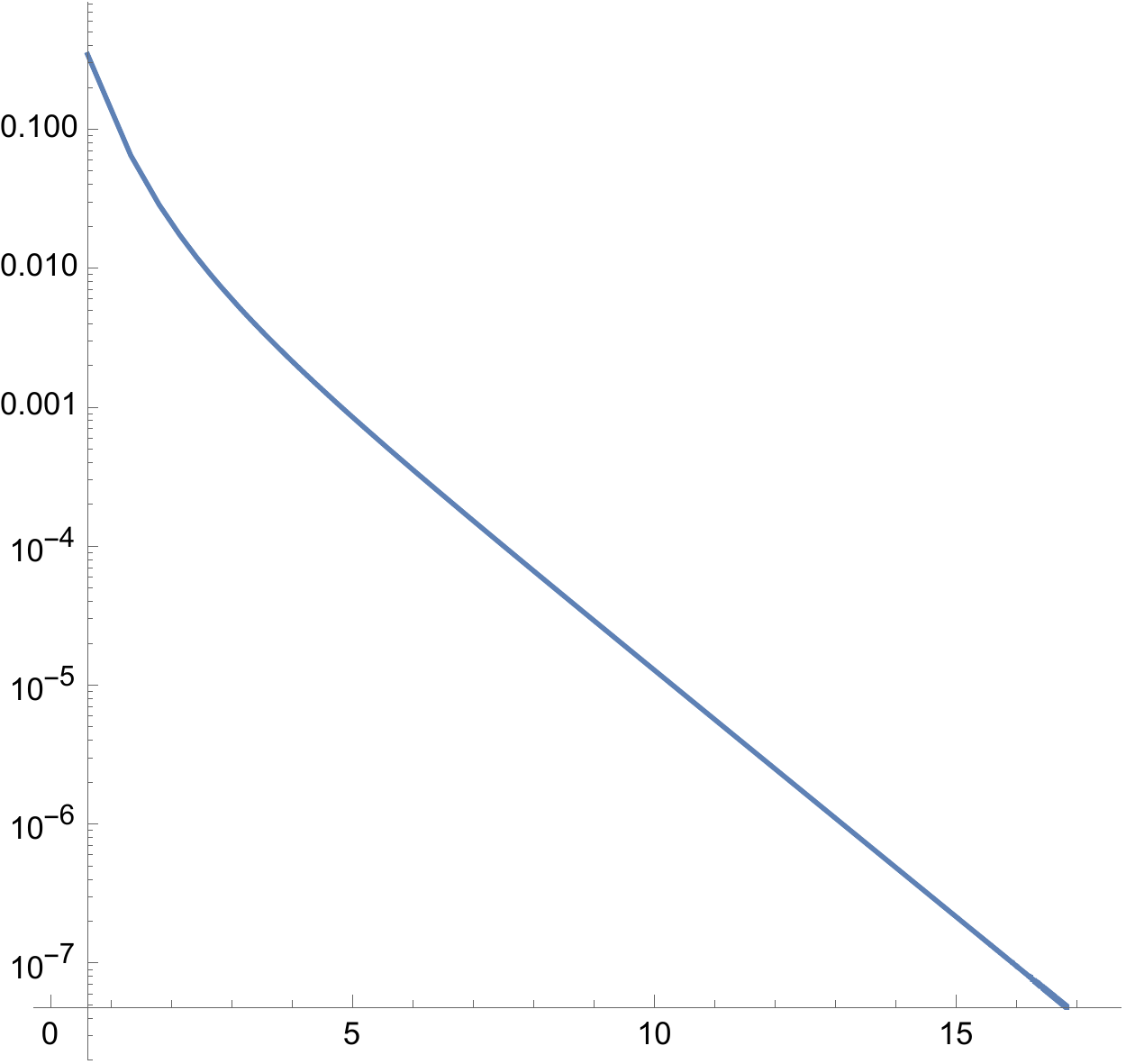}
\end{center}
\begin{picture}(0,0)
\put(230,36){$\phi$}
\put(22,237){$\Lambda^2_s$}
\end{picture}
\vspace*{-40pt}\caption{Large volume phase.\label{fig:LVphase}}
\end{subfigure}
\caption{\label{fig:Quintic} Plot of the square of the species scale for the quintic against the distance $\phi$ in scalar field space along \eqref{eq:quinticgeodesic}. In \ref{fig:LGphase} we considered the LG phase, with the right-end corresponding to its end at $|x|=1$; in \ref{fig:LVphase} the large volume phase starting from this edge $|x|=1$.}
\end{figure}
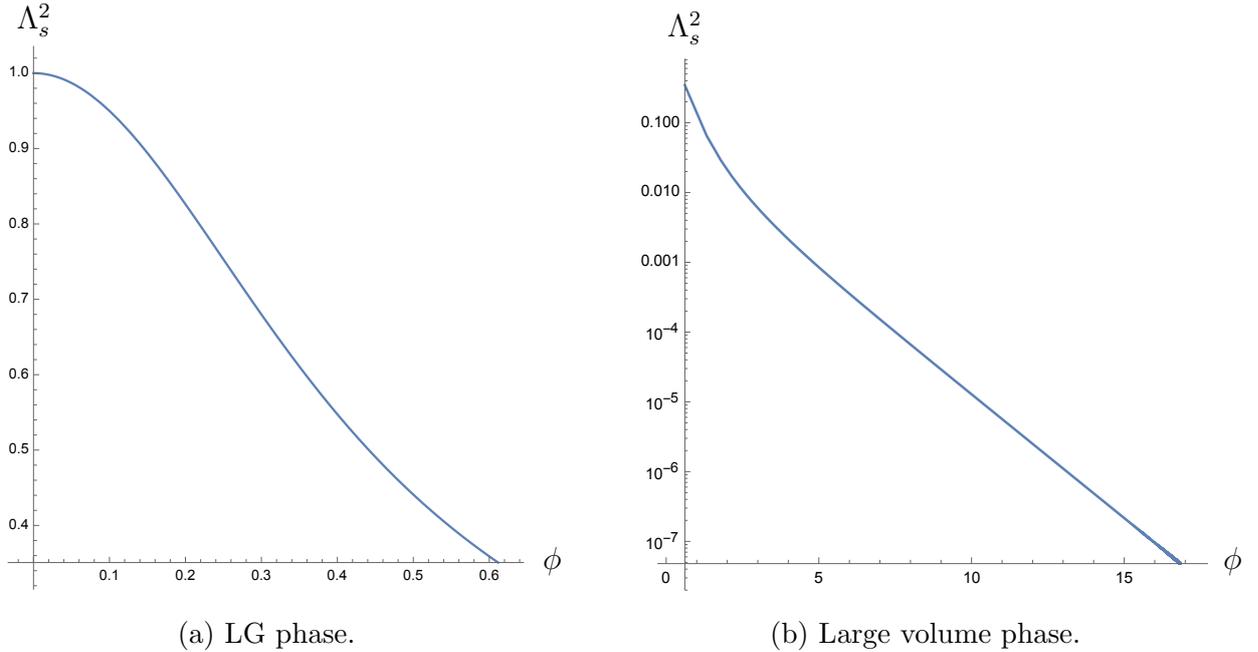

Next we consider scalar potentials over the one-dimensional K\"ahler moduli space of the mirror quintic. We parametrize this moduli space by a coordinate $x$; it has an infinite distance limit corresponding to the large volume point(s) at $x=\infty$, conifold point(s) at $x=1$, and a Landau-Ginzburg point at $x=0$. As studied in \cite{vandeHeisteeg:2022btw}, the desert point is located at the LG-point. We take this desert point as the starting point for computing the range of the potentials, and move the other endpoint out towards the large volume point. To be precise, we consider the geodesics parameterized by
\begin{equation}\label{eq:quinticgeodesic}
    x(s) = e^{\pi i/5} s \, , \qquad 0 \leq s < s_{\rm LCS}\, ,
\end{equation}
avoiding the conifold singularities.  The endpoint $s_{\rm LCS}$ is defined as the point where the species scale $\Lambda_s^2$ crosses the scalar potential $V_0$. We have provided two plots for the species scale as a function of distance along this path for both the non-geometric phase close to the LG point and as we move towards the large volume point in figure \ref{fig:Quintic}.

We next want to provide a numerical relation between the field range $\Delta \phi$ and the gap of the scalar potential $V_0$ in the large volume phase. To this end we can use the asymptotics of the species scale as
\begin{equation}\label{eq:quinticapprox}
    \Lambda_s^2 = \frac{3}{25\pi s}+\mathcal{O}(s^{-2}\log[s])\, ,
\end{equation}
where $s$ is the K\"ahler parameter. For the field range $\Delta \phi$ we can compute the distance in both the LG phase and the large volume phase along the geodesic \eqref{eq:quinticgeodesic}, where we take the endpoint to be given by where $\Lambda_s^2$ in \eqref{eq:quinticapprox} crosses $V_0$. Altogether this yields
\begin{equation}\label{eq:deltaphiquintic}
    \Delta\phi \simeq -\sqrt{\frac{3}{2}} \log[V_0] - 3.798\, ,
\end{equation}
where the constant $-3.79$ indeed matches with the plot in figure \ref{fig:LVphase}. The coefficient $\sqrt{3/2}$ matches with the expectation for a decompactification limit from four to five dimensions, in which case the species scale bound scales as
\begin{equation}
V_0 \lesssim \Lambda^2_s \sim e^{-\frac{2}{\sqrt{6}} d} \, .
\end{equation}

\subsubsection{Example 3: Mirror Bicubic \texorpdfstring{$X_{3,3}(1^6)$}{X33}}
For our next example we consider the bicubic: its moduli space contains a K-point, conifold point and a large volume point, located at $x=0$, $x=x_c=3^{-9}$ and $x=\infty$, respectively. Note that there is no orbifold point in this moduli space, so this allows us to test our bound \eqref{lambdavsV} in an example with a desert point at a generic point in moduli space; in \cite{vandeHeisteeg:2022btw} it was found to be located on the real line between the conifold point and K-point.

For our geodesic we consider a path along the real line $x>0$ that crosses through the desert and conifold point, with its endpoints in the asymptotic regimes of the large volume and K-point, given by
\begin{equation}
    x(s) = s \, , \qquad s_K \leq s \leq s_{LV}\, ,
\end{equation}
where the endpoints $s_K$ and $s_{LV}$ correspond to where the species scale $\Lambda_s^2$ crosses the gap of the scalar potential $V=V_0$. 

Let us now compute the field range of this constant scalar potential by breaking up the distance as described by \eqref{DeltaphiV0}
into two asymptotic segments and a bulk piece. Assuming $V_0$ is sufficiently small we traverse through the entire conifold phase; we compute the diameter of this interior phase between the two half-points $x=x_c/2$ and $x=3x_c/2$ to be
\begin{equation}
    \Delta\phi_{\rm bulk} \simeq 0.210\, .
\end{equation}
In turn, we consider the large volume and K-point phase separately. For both we compute the distance from the respective half-points $x_c/2$ and $3x_c/2$ up to the points $s_{LV}$ and $s_{K}$ where $V_0= \Lambda_s^2$. This yields functional expressions for the distances as a function of the gap $V_0$, where for the LCS phase we find
\begin{equation}
    \Delta\phi_{LV} \simeq -\sqrt{\frac{3}{2}} \log[V_0]-4.334\, .
\end{equation}
while for the K-point we find
\begin{equation}
    \Delta\phi_{K} \simeq -\frac{1}{\sqrt{2}}\log[V_0]-0.692\, .
\end{equation}
Putting all these three distances together we find
\begin{equation}
    \Delta\phi = \Delta\phi_{\rm bulk}+\Delta\phi_{LV}+\Delta\phi_K \simeq -\frac{1+\sqrt{3}}{\sqrt{2}} \log[V_0]-4.815\, .
\end{equation}
Note that this yields a larger field range for the potential compared to the quintic in \eqref{eq:deltaphiquintic}, as we have an additional $\log[V_0]/\sqrt{2}$ coming from moving the other endpoint to the K-point (instead of fixing it at the desert point).

\section{Concluding Remarks}\label{sec:conclusions}
In this paper we obtained a bound on the range of $V$ for slowly varying fields using the fact that in a consistent EFT, $V<\Lambda_s^2$.  Moreover, using bounds on the asymptotic form of $\Lambda_s$ coming from the distance conjecture and the emergent string conjecture, we found that our result is consistent with an asymptotic bound that was obtained from a completely different reasoning namely by using the TCC \cite{Bedroya:2019snp}. Our result in this paper refines this result in that it leads to a computable bound everywhere, not just asymptotically, and allows in the asymptotic regions to find the subleading $\mathcal{O}(1)$ shifts in the bound.  Given the fact that the nature of our argument is completely distinct from reasoning behind the TCC, it lends further support to the TCC itself.

It is natural to wonder what implications our results have for inflationary models, as having a large field range is a desirable feature in many inflation models. In these models the field range is given in terms of the Lyth bound (setting, again, the reduced Planck scale $M_{\rm pl}=1$)~\cite{Lyth:1996im}
$$\Delta \phi=\sqrt{2 \epsilon} \ N\,,$$
where $N$ is the number of e-folds and $\epsilon$ is the slow roll parameter $\epsilon={1\over 2}(V'/V)^2$.
The condition for inflation to address the horizon problem the number, $N$, of e-fold expansion during inflation should satisfy
$$e^N\geq \bigg({V\over \Lambda}\bigg)^{1/4}\,,$$
where $V$ is the inflation potential and $\Lambda\sim 0.7\times 10^{-120}$ is the dark energy
in reduced Planck units (where here we are using the coincidence of  matter-radiation equality temperature with the onset of dark energy dominance).
Using the bound in this paper ($\Delta \phi \lesssim \sqrt{6} \ \log(1/V)$), and requiring $\epsilon$ to be a small number we find (in reduced Planck units)
$$ V\lesssim 5\times 10^{-18 \sqrt{\epsilon}}\, ,$$
which is not a strong bound. This bound is very sensitive to the largest possible value for the coefficient of the $\log(1/V)$-term in $\Delta \phi$, which we fixed here using the emergent string conjecture. Notice that inflation in the context of TCC \cite{Bedroya:2019tba} puts much stronger restrictions on $V$ and $\epsilon$, since TCC leads to an upper bound on the scale of inflation:
$$e^N<{1\over H_{\rm inf}}\rightarrow \bigg({V\over \Lambda}\bigg)^{1/4}<{1\over \sqrt V}\rightarrow V^{1/4}<\Lambda^{1/12}\, .$$
Interestingly enough this scale $\Lambda^{1/12}={\hat M}_{\rm pl}\sim 10^{10}\,{\rm GeV}$ is
the Planck scale in the dark dimension scenario, which predicts one extra dimension
in our universe \cite{Montero:2022prj} in the micron range!  In that context the smallness of $V$ requires no fine tuning, namely it states:
$$V^{1/4}<\Lambda^{1/12}\rightarrow \Lambda_{\rm inf}<{\hat M}_{\rm pl}\, .$$
Moreover to get the observed power spectrum for fluctuations one would need 
$$\bigg({V\over \epsilon}\bigg)\sim 10^{-9}\, ,$$
which leads to $\epsilon \sim 10^9 V\sim 10^9 \Lambda^{1/3}\sim 10^{-30}$ for the upper range of $V$.  Thus the bound we get for the $V$ is well satisfied with such a small $\epsilon$.  Equivalently, the Lyth bound implies that
the field range for the inflaton field is related to $N$ by an extra factor of $\sqrt \epsilon$ leading to
$$\Delta \phi \sim 10^{-15} M_{\rm pl}\sim 10^{-5}{\hat M_{\rm pl}}\, .$$
which is far smaller than the upper bound derived in this work.  This small value for the potential and field range is a strong fine tuning from the 4d perspective. However from the view point of the dark dimension scenario \cite{Montero:2022prj} all we are led to is that the inflation scale is no higher than the 5d Planck scale. The inflationary period then starts near the top of the potential where it is flat (as $\epsilon\ll 1$) and the required field range is still a bit fine tuned to $10^{-5}$, but not as much as before.  So the dark dimension scenario somewhat reduces, but does not eliminate the tension, between inflation and the Swampland conditions.\footnote{A more interesting possibility in the context of inflation in the dark dimension scenario has been suggested in \cite{Anchordoqui:2022svl} involving a five-dimensional inflation.  This is also natural from the viewpoint of the present paper because, as we have found here, one extra dimension opening up leads to the largest available range for the field range for a fixed positive potential.   In this context it is natural to view the inflation
happening in the five-dimensional spacetime which could be the reason for a large and smooth fifth dimension.  Indeed the number of e-folds being the right number automatically guarantees that if the inflation potential is of the order of the 5d Planck scale, the 5d Planck length inflates to the size of the fifth dimension $\hat l_p\rightarrow l_5$ at the end of inflation because
$$e^N\sim \bigg({V\over \Lambda} \bigg)^{1/4}\sim \bigg({l_5\over {\hat l_p}}\bigg)\, .$$}

\subsubsection*{Acknowledgments} 
We would like to thank L. Anchordoqui, I. Antoniadis, A. Bedroya, D. L\"ust, and I. Valenzuela for interesting discussions and correspondence. The work of CV, MW and DW is supported by a grant from the Simons Foundation (602883,CV), the DellaPietra Foundation, and by the NSF grant PHY-2013858.

\bibliography{papers_Max}
\bibliographystyle{JHEP}

\end{document}